\documentclass[11pt,twoside]{article}

\usepackage{asp2006}
\usepackage{epsf}
\usepackage{psfig}
\usepackage{lscape}
\usepackage{graphicx}

\begin{document}
\title{The Deepest Image of the Universe at a\\ Wavelength of 15 microns}   
\author{Rosalind Hopwood,$^1$ Stephen Serjeant,$^1$ Mattia Negrello,$^1$ Chris Pearson,$^{1,2,3}$ Eiichi Egami,$^4$ Myungshin Im,$^5$ Jean-Paul Kneib,$^6$ Jongwan Ko,$^5$ Ian Smail$^7$}   
\affil{
$^1$ Department of Physics \& Astronomy, The Open University, UK\\
$^2$ Space Science \& Technology Department, CCLRC Rutherford Appleton Laboratory, UK\\
$^3$ Department of Physics, University of Lethbridge, Canada\\
$^4$ Department of Astronomy, The University of Arizona, USA\\
$^5$ Department of Physics \& Astronomy, FPRD, Seoul National University, Korea\\
$^6$ OAMP, Laboratoire d'Astrophysique de Marseille, France\\
$^7$ Institute for Computational Cosmology, Durham University, UK
}    

\begin{abstract} 
We present photometry, photometric redshifts  and extra galactic number counts for ultra deep 15 micron mapping of the gravitational lensing cluster Abell 2218 (A2218), which is the deepest image taken by any facility at this wavelength. This data resolves the cosmic infrared background (CIRB) beyond the 80\% that blank field \textit{AKARI} surveys aim to achieve. To gain an understanding of galaxy formation and evolution over the age of the Universe a necessary step is to fully resolve the CIRB, which represents the dust-shrouded cosmic star formation history. Observing through A2218 gives magnifications of up to a factor of 10, thus allowing the sampling of a more representative spread of high redshift galaxies, which comprise the bulk of the CIRB. 19 pointed observations were taken by \textit{AKARI}'s IRC MIR-L channel, and a final combined image with an area of 122.3 square arcminutes and effective integration time of 8460 seconds was achieved. The 5$\sigma$ sensitivity limit is estimated at  41.7~$\mu$Jy. An initial 5$\sigma$ catalogue of 565 sources was extracted giving 39 beams per source, which shows the image is confusion limited. Our 15 micron number counts show strong evolution consistent with galaxy evolution models that incorporate downsizing in star formation.
\end{abstract}



\section{Data Reduction}   
Our data consists of 19 pointed observation of A2218, taken with the 15 micron filter of the IRC-L \citep{on2007} aboard \textit{AKARI} \citep{mu2007}. The data was reduced using the standard IRC pipeline, version 20070912. The pipeline's median sky subtraction was utilized but resulted in dark areas, significantly around the brighter sources. To deal with this issue, and the remnants of scattered light persisting post-pipeline, a further median sky subtraction of the background areas was performed. Hot pixels were masked and removed, and the remaining bad pixels were addressed using an IDL sigma filtering routine. Figure~1 shows four corresponding postage stamp sections taken from frame 19 illustrating the post-pipeline output, a median filtered mask, the median subtracted result and the sigma filtered result. The resulting 19 images were aligned with Aladin and IDL's HASTROM, and then average combined to give the final L15 image.

\begin{figure}[!ht]
\begin{center}
\includegraphics[width=0.5\textwidth]{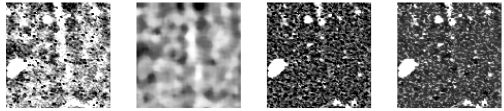}
\end{center}
\caption{
Postage stamp image sections showing, from left to right, an example of the post pipeline output, median filtered mask, median subtracted image and sigma filtered image.
}\label{fig:reductions}
\end{figure}

\subsection{Source Extraction and Completeness}   
A 5$\sigma$ source extraction was performed with DAOFIND \citep{st1987}. The resulting catalogue was eyeballed to eliminate any spurious detections, giving a final number of 565 sources. A Monte Carlo completeness test was performed using an IDL routine written to convolve a normalized empirical PSF with the final L15 image, to create artificial sources at random positions. The artificial sources were placed sufficiently apart from one another and known sources to avoid self-confusion. The test was performed for 80 flux bins, covering the range of flux densities for detected sources within the L15 image. In order to reduce statistical errors the test was repeated until an effective 18452 sources per bin was achieved. The completeness test results show the L15 image is 50\% complete to 30.7~$\mu$Jy and 80\% complete to 39.4~$\mu$Jy.

\subsection{Sensitivity}   
Aperture photometry was taken at random positions on the final image, excluding the edges and the brightest source, and the results were plotted as a histogram of flux density against number, see Figure~2. The resulting distribution has an asymmetric tail that signifies the contributions from bright sources. The combined confusion and detector noise can be represented by fitting a Gaussian to the histogram, with a standard deviation of 8.33~$\mu$Jy giving a 5$\sigma$ sensitivity estimate of 41.67~$\mu$Jy. The mean rms per pixel of the L15 noise map is 2.20~$\mu$Jy.

\begin{figure}[!ht]
\begin{center}
\includegraphics[width=0.40\textwidth]{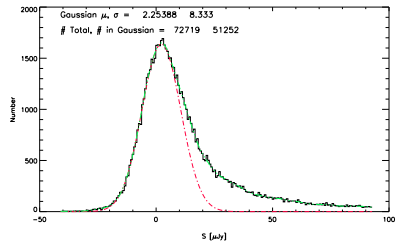}
\end{center}
\caption{
Distribution of random aperture photometry of the L15 image, fitted with a Gaussian of standard deviation 8.33~$\mu$Jy. 
}\label{fig:sensitivity}
\end{figure}

\section{Band-merged catalogue}   
Multi waveband data of A2218 was used to identify counterparts of the L15 5$\sigma$ source catalogue. HST WFP2 F450, F606 and F814, Palomar 200inch Hale \textit{U, v, b, i} and INGRID WFC \textit{Ks} and \textit{J} images of A2218 were provided by Ian Smail. Spitzer IRAC Ch 1 to 4 data was obtained via Leopard and combined with Mopex. An \textit{AKARI} S11 image was provided by Myungshin Im and Jongwan Ko, and a Spitzer MIPS 24~$\mu$m image was provided by Eiichi Egami. The counterparting procedure identified an extra 368 sources and 3 spurious 5$\sigma$ detections, giving a combined counterpart catalogue of 930 sources.

\section{Photometry}   
Aperture photometry of the 5$\sigma$ source catalogue was taken with PHOT \citep{st1987} using an aperture of 5.96$^{\prime\prime}$ and a sky annulus of radii 19.07$^{\prime\prime}$ and 31.0$^{\prime\prime}$. An aperture correction of 1.30, derived using a growth curve correction method, was applied and the IRC data user's manual version 1.4 \citep{lo2008} conversion factor of 1.691 was used to convert from ADU to~$\mu$Jy. For the HST images, photometry was obtained from the published catalogue of \citet{sm2001} via the NED database. The remaining images were subject to a growth curve correction method to obtain corrected aperture photometry, using a routine written in IDL. For each image one or two mean growth curves were empirically constructed. These curves were used to calculate aperture radius and correction for each remaining source. IDL's APER was used to take the subsequent aperture photometry. Our \textit{Ks} band photometry was compared, where available, to previously published \textit{Ks} band photometry \citet{sm2001} and showed a less than 2\% difference.

\section{Photometric Redshifts}   
We used EaZy \citep{bvdc2008} to gain photometric redshift estimates for our 5$\sigma$ source catalogue. EaZy utilizes a minimum $\chi$$^2$ SED fitting method, which is suitable for data sets with no available spectroscopic redshifts (Zspec) or a biased set of Zspec. In our case the majority of the small Zspec available are biased at the cluster distance. The theoretical SED templates, used by EaZy, are based on semi-analytical models, and a linear combination of templates can be fitted simultaneously. EaZy gives the option to apply priors, aimed at breaking the template colour degeneracies seen with increasing redshift. A comparison of the resulting redshift estimates for our sources with known Zspec shows that applying priors gives an improved correlation of approximately 10\%. Our spectra were also fitted using the photometric redshift code illustrated in \citet{ne2009}, Photz from here on. This code is uniquely optimised for fitting mid-to-far-infrared PAH and silicate features seen in starburst SEDs. Starburst template \citep{tk2003} and AGN template \citep{err1995} components were simultaneously fitted by photz. A comparison of the Photz and EaZy redshifts estimates for sources with prominent mid-to-far-infrared features shows a correlation of around 0.8.

\section{Number Counts}   
Differential number counts (dN/dS), corrected for incompleteness, were taken for our 5$\sigma$ catalogue and normalized to a  Euclidean slope. These counts were corrected for flux amplification by applying magnification factors obtained via LENSTOOL \citep{ju2007}. Figure~3 compares our corrected and uncorrected counts with previously published differential number counts. Our counts corrected for lensing show an upturn around 2 mJy and peak around 0.4 mJy, in agreement with previous counts \citep[e.g.,][]{ez1999} and the \citet{CP2007,CP2009} model.

\begin{figure}[!ht]
\begin{center}
\includegraphics[width=0.71\textwidth]{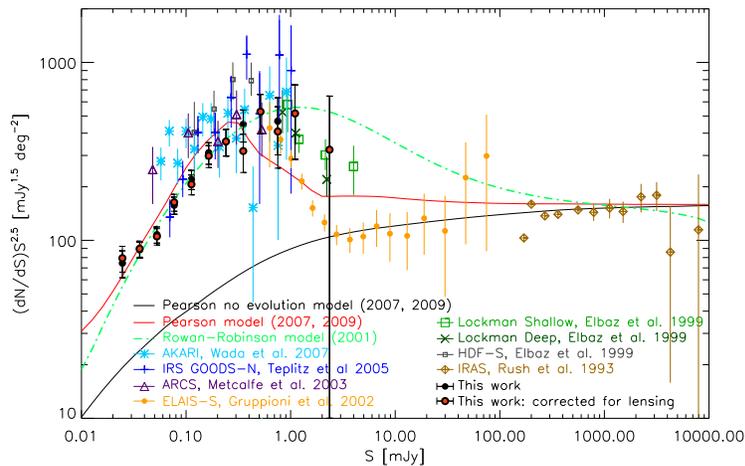}
\end{center}
\caption{
Euclidean normalized differential number count comparison.
}\label{fig:counts}
\end{figure}

\acknowledgements We thank the Great Britain Sasakawa Foundation for support with grant number 3108. This research is based on observations with \textit{AKARI}, a JAXA project with the participation of ESA.



\begin{thebibliography}{}
\bibitem[Brammer, van Dokkum \& Coppi(2008)]{bvdc2008}
Brammer, R. G., van Dokkum, P. G., \& Coppi, P. 2008, ApJ, 686, 1503
\bibitem[Efstathiou \& Rowan-Robinson(1995)]{err1995}
Efstathiou, A., \& Rowan-Robinson, M. 1995, MNRAS, 273, 649
\bibitem[Elbaz et al.(1999)]{ez1999}
Elbaz, D., Cesarsky, C. J., Fadda, D., et al. 1999, A\&A, 351, L37
\bibitem[Jullo et al.(2007)]{ju2007}
Jullo, E., Kneib, J.-P., Limousin, M., El\'{i}asd\'{o}ttir, \'{A}., Marshall, P., \& Verdugo, T. 2007, NJPh, 9, 447
\bibitem[Lorente et al.(2008)]{lo2008}
Lorente, R., Onaka, T., Ita, Y., Ohyama, Y., Tanab\'{e}, T., Pearson, C. P.  2008, AKARI IRC Data User Manual Version 1.4
\bibitem[Murakami et al.(2007)]{mu2007}
Murakami, H., et al. 2007, PASJ, 59, S369,
\bibitem[Negrello et al.(2009)]{ne2009}
Negrello, M., et al. 2009, MNRAS, 394, 375
\bibitem[Onaka et al.(2007)]{on2007}
Onaka, T. et al. 2007, PASJ, 59, S401
\bibitem[Pearson et al.(2007)]{CP2007}
Pearson, C. P., et al. 2007, Advances in Space Research, 40, 605
\bibitem[Pearson, C. (2009)]{CP2009}
Pearson, C. 2009, in preparation
\bibitem[Smail et al.(2001)]{sm2001}
Smail, I., Kuntschner, H., Kodama, T., et al. 2001, MNRAS 323, 839
\bibitem[Stetson(1987)]{st1987}
Stetson, P. B., 1987 PASP, 99, 191
\bibitem[Takagi et al.(2003)]{tk2003}
Takagi, T., Arimoto, N., Hanami, H. 2003, MNRAS, 340, 813
\end{thebibliography}
\end{document}